\documentclass[
aps,prl,
reprint,
a4paper,
superscriptaddress,
floatfix,
]{revtex4-1}
\usepackage[utf8]{inputenc}
\usepackage[T1]{fontenc}

\usepackage{amsmath,amsthm,amsfonts}
\usepackage{braket}
\usepackage{graphicx}
\usepackage{rotating}
\usepackage{hyperref}
\usepackage{units}
\usepackage{xfrac}

\hypersetup{citecolor=magenta}
\hypersetup{colorlinks=true}
\hypersetup{linkcolor=blue}
\hypersetup{urlcolor=blue}

\DeclareMathOperator{\argmin}{arg\,min}
\DeclareMathOperator{\im}{Im}

\DeclareMathOperator{\tr}{tr}
\newcommand{\N}{{\phantom{0}}}
\newcommand{\abs}[1]{{\lvert{#1}\rvert}}
\renewcommand{\doi}[2]{{\href{http://doi.org/#1}{#2}}}
\newcommand{\ee}{{\mathrm e}}
\newcommand{\ketbra}[2]{{\ket{#1}\!\!\bra{#2}}}
\newcommand{\proj}[1]{{\ketbra{#1}{#1}}}

\begin{document}


\iftrue
\title{Tracking the dynamics of an ideal quantum measurement}
\else
\title{Tracking the dynamics of an ideal quantum measurement in a trapped-ion 
 fluorescence experiment}
\fi

\author{Fabian Pokorny}
\email{fabian.pokorny@fysik.su.se}
\affiliation{Department of Physics,
Stockholm University,
10691 Stockholm, Sweden}

\author{Chi Zhang}
\email{chi.zhang@fysik.su.se}
\affiliation{Department of Physics,
Stockholm University,
10691 Stockholm, Sweden}

\author{Gerard Higgins}
\email{gerard.higgins@fysik.su.se}
\affiliation{Department of Physics,
Stockholm University,
10691 Stockholm, Sweden}

\author{Adán Cabello}
\email{adan@us.es}
\affiliation{Departamento de Física Aplicada II,
Universidad de Sevilla,
E-41012 Sevilla, Spain}

\author{Matthias Kleinmann}
\email{matthias.kleinmann@uni-siegen.de}
\affiliation{Department of Theoretical Physics,
University of the Basque Country UPV/EHU,
P.O.\ Box 644, E-48080 Bilbao, Spain}
\affiliation{Naturwissenschaftlich--Technische Fakultät,
Universität Siegen,
Walter-Flex-Straße 3, 57068 Siegen, Germany}

\author{Markus Hennrich}
\email{markus.hennrich@fysik.su.se}
\affiliation{Department of Physics,
Stockholm University,
10691 Stockholm, Sweden}

\begin{abstract}
The existence of ideal quantum measurements is one of the fundamental predictions of quantum mechanics.
In theory the measurement projects onto the eigenbasis of the measurement observable while preserving all coherences of degenerate eigenstates.
The question arises whether there are dynamical processes in nature that correspond to such ideal quantum measurements.
Here we address this question and present experimental results monitoring the dynamics of a naturally occurring measurement process: the coupling of a trapped ion qutrit to the photon environment.
By taking tomographic snapshots during the detection process, we show with an average fidelity of $94\%$ that the process develops in agreement with the model of an ideal quantum measurement.
\end{abstract}

\maketitle

{\em Introduction.---}%
What is an ideal measurement in quantum mechanics?
What are its inner workings?
How does the quantum state change because of it?
These have been central questions in the development of quantum mechanics 
 \cite{WZ83}.
Notably, today's accepted answer to the latter question is conceptually 
 different from the one given in the first formalization of quantum mechanics 
 by von Neumann \cite{vonNeumann32}.
Then, it was thought that an ideal measurement on a quantum system would 
 inevitably destroy all quantum superpositions.
Later, Lüders pointed out \cite{Luders51} that certain superpositions should 
 survive, so that a sequence of ideal measurements would preserve quantum 
 coherence.
Lüders version is the one accepted today.

Even though there is agreement on the theoretical description of an ideal 
 measurement, it is a fundamental question whether, and if so how, ideal 
 measurements occur as natural processes.
This is an especially sensitive question as measurements do not fall into the 
 domain of unitary time evolution.
In this Letter, we demonstrate a natural dynamical process that realizes an 
 ideal quantum measurement.
For this, we implement a natural process that is considered to be an ideal 
 measurement, and monitor its dynamics by taking a sequence of snapshots while 
 the process is occurring.
These snapshots are tomographically complete and allow us to compare the 
 experimental results with the theoretical prediction of an ideal measurement.
 
Ideal measurements model an ideal implementation of a quantum observable.
In the discrete case, a measurement of the observable $A$ yields values $a_k$ 
 according to the spectral decomposition $A= \sum_k a_k \Pi_k$, where $\Pi_k$ 
 are mutually orthogonal projections summing to identity, and $a_k$ are distinct real 
 numbers.
Any measurement requires an interaction of the measurement apparatus with the 
 system and this interaction affects the state of the system.
This is true, independently of whether or not the measurement result is 
 recorded by an observer.
However, the effect on the state depends on the experimental realization.
In practice, a measurement is often a rather violent process that ``destroys'' 
 the quantum system, for example the detection of a photon. 

A Lüders process is the ideal implementation of a measurement in which the 
 state of the system is transformed according to \cite{Heinosaari12}
\begin{equation}
 \Xi_A\colon \rho\mapsto \sum_k \Pi_k \rho \Pi_k.
\end{equation}
(For simplicity, we assume that the observer ignores the measurement outcome.)
From a theoretical perspective, this process is truly special.
On the one hand, it is the only process implementing a measurement of $A$ which 
 does not disturb any subsequent refined measurement $B$ \cite{Chiribella14, Kleinmann14}.
That is, a measurement $B$ where each outcome $\ell$ is at most as likely as a 
 certain outcome $k_{\ell}$ of $A$, for all states $\rho$.
This implies that, whenever two observables $A$ and $B$ are compatible, $AB= 
 BA$, then the respective Lüders processes do not disturb each other, 
 $\Xi_A\Xi_B= \Xi_B\Xi_A$ \cite{Luders51, Heinosaari12}.
On the other hand, $\Xi_A$ is universal: any process $\Lambda_A$ describing a 
 measurement of $A$ can be written as $\Lambda_A\colon \rho \mapsto 
 \sum_k\Phi_k(\Pi_k\rho\Pi_k)$, where $\Phi_k$ are some probability-preserving 
 processes \cite{Heinosaari12}.

While these properties constitute strong theoretical arguments for the special 
 role of the Lüders process, this does not imply that Lüders processes occur in nature.
An important aspect of the Lüders process is that it leaves any quantum 
 superposition of degenerate eigenstates unaffected.
In this sense, the existence of Lüders processes is a nontrivial prediction of 
 quantum mechanics that is usually taken for granted rather than tested 
 thoroughly.

In order to investigate whether Lüders processes do exist in nature, we 
 consider the canonical model for measurements \cite{Peres95, Heinosaari12}:
A system in state $\rho$ is brought into contact with a pointer 
 system, which is in state $\ket{\Phi}$.
When in contact, the two systems interact via the Hamiltonian $H_A$ for a time 
 $\tau$ and, after separating the two systems, the pointer system is measured 
 in some fixed basis $\ket{\omega_j}$.
This yields the process
\begin{equation}\label{eq:pointer}
 M_A\colon \rho\mapsto
 \sum_j \bra{\omega_j}\ee^{-i H_A\tau/\hbar} \left(\rho\otimes \proj\Phi\right)
 \ee^{i H_A\tau/\hbar}\ket{\omega_j}.
\end{equation}
However, the class of processes covered by Eq.~\eqref{eq:pointer} is so 
 general, that it can be shown \cite{Heinosaari12} that any quantum process, 
 including the Lüders processes $\Xi_A$, can be modeled by 
 Eq.~\eqref{eq:pointer}.
Indeed, if we are able to engineer interactions and pointer systems at will as 
 in, for example, quantum processors \cite{Deutsch85, Barenco95}, the 
 fact that a process cannot be approximated by Eq.~\eqref{eq:pointer} would be 
 in contradiction of our present understanding of those systems.
Usually, experimental measurement procedures are not Lüders processes, because 
 this would require a careful fine-tuning of the Hamiltonian $H_A$ or else the degeneracy of the observable would be lifted by experimental 
 imperfections which then leads to a process as the one envisioned by von 
 Neumann, in which all coherences are destroyed.

{\em Fluorescence measurements.---}%
The best candidate for a natural Lüders process is a interaction-free measurement, that 
 is, a measurement of an observable of the form $A= \proj{\phi}$, where the 
 event ``detection'' corresponds to the eigenvalue $1$ and the event ``no 
 detection'' corresponds to the eigenvalue $0$.
There, it can be expected that the event ``no detection'' does not require any 
 interaction with the eigenspace with eigenvalue $0$ and hence superpositions 
 in this subspace are preserved.
Our choice for an experiment within which to identify a Lüders process is the coupling
 of a single trapped ion to the photon environment.
To explain why this is a good candidate, it is useful to present a simplified 
 description of the specific physical process in our experiment.
 For a more 
 accurate theoretical model of this process see Ref.~\cite{Carmichael93} and 
 the Appendix.

We prepare an ion in a superposition of three electronic states, 
 $\ket0$, $\ket1$, and $\ket2$.
By driving the transition $\ket{0}$ to a short-lived 
 excited level $\ket e$, we facilitate the emission of photons into the 
 environment, via the natural decay $\ket{e}\ket{n=0}\rightarrow \ket 
 0\ket{n=1}$.
Here, $\ket{n=0}$ is the initial state of the photon environment and 
 $\ket{n=1}$ the state of the photon environment after a photon 
 has been emitted by the decay.
We write the initial state of the system and the photon environment as
\begin{equation}
 \ket\Psi= (\alpha_0\ket0+ \alpha_1\ket1+ \alpha_2\ket2) \ket{n=0}.
\end{equation}
Since only the level $\ket0$ participates in the driving $\ket0\leftrightarrow 
 \ket{e}$ and the subsequent decay, the state after the driving is
\begin{equation}\begin{split}
 \ket{\Psi_m}= \alpha_0\ket0 \left( g_0\ket{n=0} + g_1 
 \ket{n=1}\right)+ \\
 (\alpha_1\ket1+ \alpha_2\ket2) \ket{n=0},
\end{split}\end{equation}
where $g_1$ is related to the probability of photon scattering by $P_{scatt} = \abs{g_1}^2$, and $\abs{g_0}^2 + \abs{g_1}^2 = 1$. 
Ignoring the photon environment, we obtain the reduced state
\begin{equation}
 \rho_m=
 \begin{pmatrix}
  \abs{\alpha_0}^2 & \alpha_0 \alpha_1^* g_0 & \alpha_0 \alpha_2^* g_0 \\
  \alpha_0^* \alpha_1 g_0^* & \abs{\alpha_1}^2 & \alpha_1 \alpha_2^* \\
  \alpha_0^*\alpha_2 g_0^* & \alpha_1^* \alpha_2 & \abs{\alpha_2}^2
 \end{pmatrix}.
\end{equation}
The coherence between the levels $\ket1$ and $\ket2$ is preserved while the 
 coherences between $\ket0$ and $\ket1$ and between $\ket0$ and $\ket2$ decay with 
 $\abs{g_0}$.

\begin{figure*}[htp]
\includegraphics[width=\linewidth]{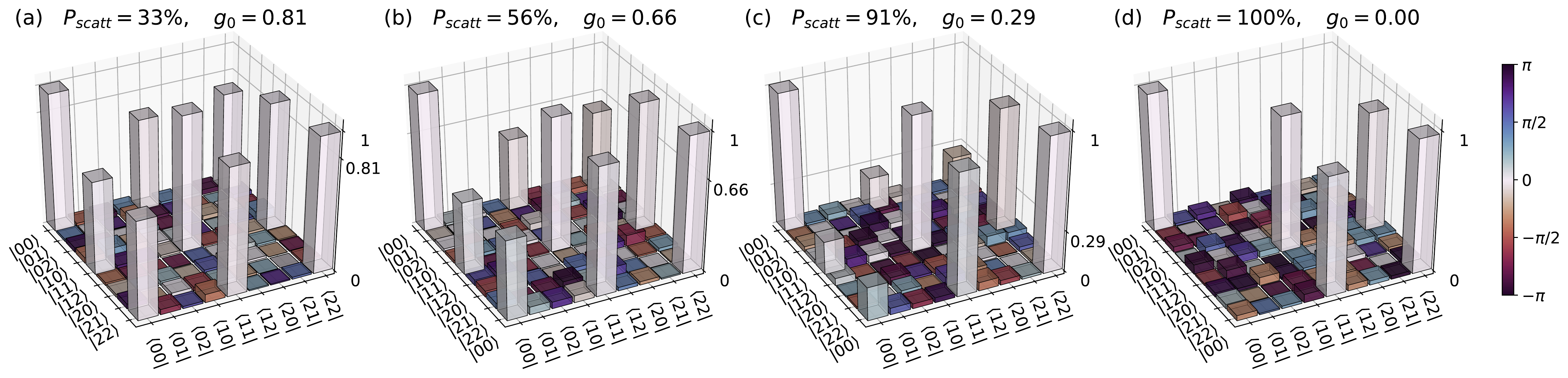}
\caption{\label{fig:chi}%
Choi matrices $\chi_\mathrm{exp}$ reconstructed from the experimental data. The colors indicate the complex phase,
$g_0$ is calculated according to our model, see Eq.~(\ref{eq:a0}).
From (a) to (d) the coupling strength of $\ket{0}$ to the photon environment is increased and coherences involving state $\ket{0}$ are destroyed.
(d) is similar to the ideal Lüders process, and from (d) $\rightarrow$ (a) the process becomes more similar to the trivial process $\rho \mapsto \rho$.}
\end{figure*}

Intuitively, if at least one photon is scattered into the environment, then 
 $\abs{g_0}= 0$ and the coupling to the photon environment corresponds to the measurement 
 of the qutrit projection $\proj 0$.
Because the probability for scattering is $P_\mathrm{scatt}= 1-\abs{g_0}^2$ 
 the implemented measurement is the generalized measurement $(E_1,E_0)$ with 
 $E_1= P_\mathrm{scatt} \proj0$ and $E_0= \openone-E_1$.
The numerical value of $g_0$ depends on the experimental configuration and its 
 computation requires a more rigorous model of the measurement process.
A formula for $g_0$ is given in Eq.~\eqref{eq:a0} below and 
 $P_{scatt}$ is varied between $0.33$ and $1.0$, see Fig.~\ref{fig:chi}.

Since our computation of the state $\rho_m$ is generic for any 
 initial state, we can read off the measurement process $\Lambda_m$ as
\begin{equation}\label{eq:lambdam}
 \Lambda_m\colon \rho\mapsto \sqrt{E_1}\rho\sqrt{E_1}
               +G\sqrt{E_0}\rho\sqrt{E_0}G^\dag,
\end{equation}
 with $G= \ee^{i \varphi} \proj 0+\proj 1+\proj 2$ and $\varphi=\arg(g_0)$.
For $P_\mathrm{scatt}\rightarrow 1$, this process implements the Lüders process 
 $\Xi_A$ for the observable $A= \proj 0$ and hence constitutes an ideal 
 measurement.
For $P_\mathrm{scatt}<1$, the process is a transitional form between $\Xi_A$ 
 and the trivial process $\rho\mapsto \rho$.
Then, not only are the coherence between $\ket1$ and $\ket2$ preserved but also 
 partially the coherences between $\ket0$ and the other two qutrit levels.
The occurrence of the terms $\sqrt{E_k}\rho\sqrt{E_k}$ in $\Lambda_m$ 
 is in accordance to the canonical form of such transitional processes 
 \cite{Heinosaari12}, while the phase $\ee^{i\varphi}$ introduced by the 
 unitary $G$ is specific to this measurement process.

\begin{figure}[htp]
\includegraphics[width=\linewidth]{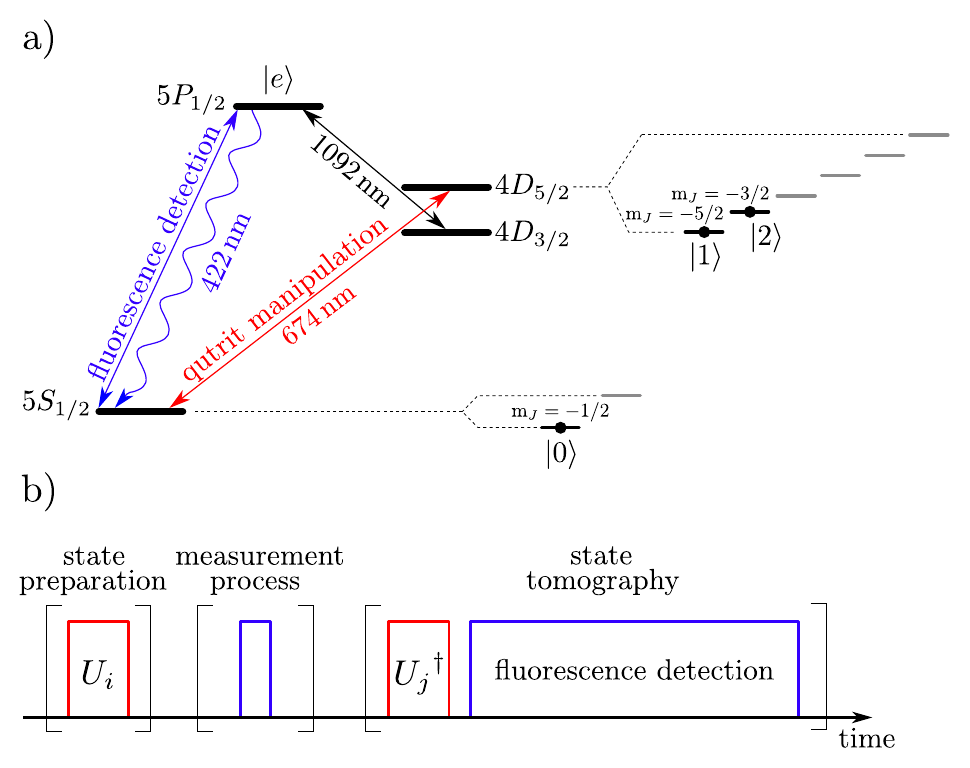}
\caption{\label{fig:experiment}
Experimental realization of the process tomography of a Lüders measurement.
(a)
Level scheme of $^{88}\mathrm{Sr}^{+}$.
A magnetic field of $\unit[(0.3576\pm 0.0009)]{mT}$ applied along the trap 
 axis splits the $\mathrm{m_J}= -5/2$ and $\mathrm{m_J}= -3/2$ Zeeman
 sublevels of the metastable states $4D_{5/2}$ by $2\pi\times \unit[6.0]{MHz}$.
Together with the $5S_{1/2}\;\mathrm{m_J}= -1/2$ ground state, these states 
 form the qutrit basis states $\ket0$, $\ket1$, and $\ket2$.
The qutrit transitions are driven by a narrow-linewidth laser at $\unit[674]{nm}$
 (linewidth $<2\pi\times \unit[600]{Hz}$) tuned to either the
 $\ket{0} \leftrightarrow \ket{1}$ or $\ket{0} \leftrightarrow \ket{2}$ transition.
A $\unit[422]{nm}$ laser driving the $\ket{0} \leftrightarrow \ket{e} \equiv 5P_{1/2}$ 
 transition is used to induce the Lüders measurement process and the fluorescence
 detection during state tomography.
To prevent loss of population from the qutrit subspace during the measurement process $\unit[1092]{nm}$ laser light and circularly-polarised $\unit[422]{nm}$ laser light is used to repump population from $4D_{3/2}$ and $5S_{1/2}\;\mathrm{m_J}= +1/2$ to $\ket{0}$.
(b)
Experimental sequence for the process tomography.
First the ion is prepared in one of nine initial states (Appendix, Table~\ref{tab:tomography}) followed 
 by a measurement which implements snapshots of an ideal 
 measurement process and finally state tomography is performed using a qutrit rotation and fluorescence detection.}
\end{figure}

{\em Experimental setup.---}%
Our setup consists of a single $^{88}\mathrm{Sr}^{+}$ ion confined in a linear 
 Paul trap, similar to Ref.~\cite{Higgins17}.
Fig.~\ref{fig:experiment}~(a) shows the level scheme of the ion with the qutrit states
 encoded in electronic states of the ion.
The measurement process is implemented using a pulse of $\unit[422]{nm}$ 
 laser light, which facilitates coupling of the electronic state $|0\rangle$
 and the photon environment, while the qutrit levels $\ket1$ and $\ket2$ are left unaffected.
We use the same laser light during the fluorescence detection step.

The variable $g_0$ is obtained from a quantum optical model of the 
 measurement, see Appendix.
This yields
\begin{equation}\label{eq:a0}
 g_0\approx \exp\left(-\frac{\Omega^2}{2\Gamma + 4 i \Delta} t\right) \cdot \exp\left(i \phi_{r} \right),
\end{equation}
 where $\Omega$ is the Rabi frequency driving the $|0\rangle \leftrightarrow |e\rangle$ transition, $\Gamma= 
 2\pi\times \unit[21.65]{MHz}$ \cite{Gallagher67} is the decay rate of state 
 $|e\rangle$, $\Delta= 2\pi\times \unit[(5\pm2)]{MHz}$ the detuning from 
 resonance, $t= \unit[1]{\mu s}$ is the length of the laser pulse and the phase $\phi_{r}$ results from the ac-Stark shifts induced by repump laser fields at $\unit[422]{nm}$ and $\unit[1092]{nm}$ which are present during the measurement process. The intensities and detunings of the repump fields are tuned such that $\im\left(g_0\right)\approx 0$.
Values of $\Omega$ and $\Delta$ are determined from the $5S_{1/2} \leftrightarrow 5P_{1/2}$ spectral lineshape.

We track the dynamics of the measurement process by carrying out
 process tomography as the power of the $\unit[422]{nm}$ laser used for coupling $|0\rangle \leftrightarrow |e\rangle$ is varied.
For the process tomography [Fig.~\ref{fig:experiment}~(b)], the ion is prepared in $\ket0$, then rotated (unitary $U_i$) to one of the nine 
 initial states $\ket{\psi_i}$ using pulses of $\unit[674]{nm}$ laser light, 
 see Table~\ref{tab:tomography} in the Appendix.
Then the measurement process is carried out; a $\unit[422]{nm}$ laser pulse is applied for $\unit[1]{\mu s}$.
Finally the state 
 tomography is carried out by applying rotation $U_j^\dag$ followed by fluorescence detection for $\unit[500]{\mu s}$.
The measurement process uses a shorter pulse length than the fluorescence detection step, since the photons scattered during the measurement process do not need to be detected.
During state tomography the rotation $U_j^\dag$ followed by a measurement of $(\proj0,\openone-\proj0)$ 
 acts as measurement operator $(\proj{\psi_j},\openone-\proj{\psi_j})$.
The nine initial states are each measured by the nine different operators; this is sufficient to characterize the experimental process.

{\em Results.---}%
For analyzing the experiment, a process $\Lambda$ is more conveniently 
 represented by its Choi matrix \cite{Heinosaari12},
\begin{equation}
 \chi= \sum_{i,j=0}^2 
\Lambda[\,\ketbra{i}j_\mathrm{sys}]\otimes\ketbra{i}j_\mathrm{aux}.
\end{equation}
This matrix is positive semidefinite for any physical process and yields a 
 complete characterization of $\Lambda$, since
 $\Lambda[\rho]= \tr_\mathrm{aux}[\chi(\openone\otimes\rho^T)]$ with
 $\rho^T= \sum_{i,j}\ketbra{i}j\braket{j|\rho|i}$.
If $\Lambda$ is a probability-preserving process, that is, $\tr(\Lambda[\rho])= 
 \tr(\rho)= 1$, then the Choi matrix obeys $\tr_\mathrm{sys}(\chi)= 
 \openone_\mathrm{aux}$.
For the measurement process $\Lambda_m$ in this work, the Choi matrix reads
\begin{equation}\label{eq:choimodel}
 \chi_m= \proj{00}+\proj{\xi}+
 g_0\ketbra{00}{\xi}+g_0^*\ketbra{\xi}{00},
\end{equation}
 where $\ket\xi= \ket{11}+\ket{22}$.

The system is prepared in $\ket{\psi_i}$, the process $\Lambda_m$ is carried out and
 subsequently the system is measured using the observable $\proj{\psi_j}$.
This is repeated $N=1000$ times for each $i,j$, with measurement outcome $1$ occuring $n_{i,j}$ times, and thus resulting in the experimental outcome frequency $f_{i,j}=n_{i,j}/N$.
We use $f_{i,j}$ to reconstruct the experimental Choi matrix $\chi_\mathrm{exp}$ with $\chi_\mathrm{exp}= \argmin_\chi \sum_{i,j} (p_{i,j}-f_{i,j})^2$
where the probabilities $p_{i,j}$ are given by
\begin{equation}\begin{split}
 p_{i,j} &= \bra{\psi_j}\Lambda[\,\proj{\psi_i}\,]\ket{\psi_j} \\
         &= \tr[\chi\, \proj{\psi_j}\otimes \proj{\psi_i}^T].
\end{split}\end{equation}
In addition we calculate the Uhlmann fidelity (rescaled by $\frac{1}{9}$) of the experimental Choi matrix and the model to be on average $94\%$.
For details see the Appendix.

$\chi_\mathrm{exp}$ does not perfectly obey the 
 probability-preserving constraint $\tr_\mathrm{sys}\chi= 
 \openone_\mathrm{aux}$, due to statistical fluctuations and imperfections in 
 the experimental setup.
A likelihood ratio test shows that the deviation from $\tr_\mathrm{sys}\chi= 
 \openone_\mathrm{aux}$ is statistically significant by between $4$ and $8$ 
 standard deviations.
This can be attributed to imperfect state preparation.
While the effect is statistically significant, it is clear from the high number 
 of samples that the systematic deviation is overall very small and can be 
 neglected.
In the Appendix (Fig.~\ref{fig:exp+data})  we show a comparison of the $\chi_{\mathrm{exp}}$ reconstructed from experimental data
 with our model predictions including the respective uncertainties.
We also show post-process density matrices for the initial state
 $\frac{1}{\sqrt{2}} \left(\ket{1} + i\ket{2}\right)$ -- the coherence is preserved, and for the initial state $\frac{1}{\sqrt{2}}\left(\ket{0} + i\ket{2}\right)$ -- the coherence is destroyed.
 
{\em Conclusion.---}%
Fluorescence detection is the standard way to measure a qubit in an ion trap or 
 in similar setups that enable quantum information processing.
This process is a prime example of an ideal measurement process:
The system which is subject to the measurement is brought into contact with a 
 macroscopic pointer state by facilitating a strong interaction between the system 
 and the photon environment.
A measurement of the photon environment then reveals the measurement outcome.
Hence the measurement should behave as predicted by quantum mechanics for 
 ideal measurements, that is, any coherence between all levels 
 that are not measured should persist.
We verified this property by 
 performing process tomography of the coupling of an electronic state of a single trapped ion to the photon environment.
The fidelity between the ideal measurement process and our implementation is 
 $94\%$ and matches the quality of a similar experiment \cite{Jerger16} in a 
 superconducting quantum system.

The theoretically ideal measurement process only occurs in the case of a strong 
 interaction.
If the interaction is weakened, then an intermediate process between the ideal 
 measurement process and the identity process occurs.
In the mathematical theory of measurements \cite{Heinosaari12}, such ideal 
 measurements have a canonical ``square root'' form, see 
 Eq.~\eqref{eq:lambdam}.
The quantum optical model for a weak measurement predicts a 
 similar form which is only different in an additional phase between $\ket0$ 
 and the other levels.
In process tomographies for those intermediate processes we obtain an overall process 
 fidelity with the predicted model of approximately $94\%$.

We thus quantitatively demonstrated how a measurement can be implemented while 
 preserving coherence and in which sense nature allows us to implement a weak 
 measurement.
A future direction of research is to test the predictions of ideal quantum 
 measurements beyond what we have tested here:
In the current experiment, the coherence-preserving measurement works 
 relatively effortless, since the measurement process only affects a 
 single state of our qutrit.
This corresponds to a interaction-free measurement where it is only measured whether or 
 not the system is in state $\ket 0$.
It could also be possible to find an ideal measurement process in nature, where 
 all eigenvalues of the measured observable are two-fold degenerate and 
 therefore the coherence between the degenerate eigenstates needs to be 
 preserved.
Whether such processes exist as natural processes and can be implemented with a 
 fidelity comparable to our experiment is an open question.

{\em Acknowledgements.---}
We thank Mohamed Bourennane and Markus Müller for discussions.
This work was supported by
the Spanish Ministry of Economy, Industry and Competitiveness MINECO and
the European Regional Development Fund FEDER through Grant No.\ FIS2017-89609-P 
 and No.\ FIS2015-67161-P,
by the FQXi Large Grant ``The Observer Observed: A Bayesian Route to the 
Reconstruction of Quantum Theory,''
by the EU (ERC Consolidator Grant No.\ 683107/TempoQ and
ERC Starting Grant No.\ 258647/GEDENTQOPT),
by the Basque Government (Grant No.\ IT986-16), by the Swedish Research Council (Trapped Rydberg Ion Quantum Simulator), by QuantERA project ``ERyQSenS'',
and by the project ``Photonic Quantum Information'' (Knut and Alice Wallenberg 
Foundation, Sweden).

\appendix
\section{Appendix: Model for measurement process}
We consider the transition between a ground state $\ket0$ and a short-lived 
 excited state $\ket e$ of the trapped ion.
This transition is driven with Rabi frequency $\Omega$
 using a classical laser field, where the laser frequency is detuned from resonance by $\Delta$.
In the rotating-wave approximation, the interaction Hamiltonian is then given 
 by \cite{Carmichael93}
\begin{equation}
 H_I= \hbar \Delta \proj{e}+ \frac{\hbar\Omega}2 (\sigma_+ + \sigma_-),
\end{equation}
 written in the frame rotating with the laser field.
Here, $\sigma_+= \ketbra{e}0= \sigma_-^\dag$.
The other levels used in our system, $\ket1$ and $\ket2$, are unaffected by the 
 driving laser, so that $H_I$ does not have contributions involving those 
 levels.

The excited state decays back to the ground state with decay rate $\Gamma$ and 
 on decay, a photon is emitted into the photonic environment.
In the Born--Markov approximation, this environment can be traced out and the 
 state $\rho$ of the levels $\ket0$, $\ket1$, $\ket2$, and $\ket{e}$ follows a 
 master equation in Lindblad form \cite{Carmichael93}.
In the interaction picture, this equation reads
\begin{equation}\label{eq:master}
 \dot{\rho}= -\frac{i}{\hbar}[H_I,\rho]+
 \frac\Gamma2(2\sigma_-\rho \sigma_+-\sigma_+ \sigma_-\rho-\rho \sigma_+ 
\sigma_-).
\end{equation}

Without the laser ($\Omega= 0$) we have $\dot \rho_{e,e}= -\Gamma \rho_{e,e}$ 
 and hence after a very short time ($\sim \Gamma^{-1}$) the excited state relaxes, 
 $\rho_{e,e}= 0$.
Since we assume that there was no population of the excited level in the 
 initial state, it is sufficient to consider the qutrit part of $\rho$.
The transformation $\Lambda_m$ induced by the measurement is then
\begin{equation}
 \Lambda_m[\rho]=
 \begin{pmatrix}
  \rho_{0,0}& \rho_{0,1}\,g_0&\rho_{0,2}\,g_0 \\
  \rho_{1,0}\,g_0^*& \rho_{1,1}&\rho_{1,2} \\
  \rho_{2,0}\,g_0^*& \rho_{2,1}&\rho_{2,2}
 \end{pmatrix},
\end{equation}
 in accordance with Eq.~\eqref{eq:lambdam}.
The variable $g_0\equiv g_0(t)$ is determined by the solution of the master 
 equation \eqref{eq:master} for $\rho_{0,1}$, namely,
\begin{align}
 \dot\rho_{0,1}&= -i\frac\Omega2 \rho_{e,1}, \\
 \dot\rho_{e,1}&= -i\frac\Omega2 \rho_{0,1}-
                 \left(\frac\Gamma2+i\Delta\right)\rho_{e,1}.
\end{align}
These equations can be solved exactly, but yield impractical expressions.
However, for $\Omega\ll \Gamma$ the excited state can be adiabatically 
 eliminated via $\dot{\rho}_{e,1}\approx 0$.
With this approximation, Eq.~\eqref{eq:a0} follows at once.

\section{Appendix: Experimental settings}
\begin{table}[h]                                                                                                                                                                                                                                                                                                                                                                                                                                                                                             
\centering
\begin{ruledtabular}
\begin{tabular}{cccccc}
 & Power [$\unit{\mu W}$] & $\Omega$ [$\unit[2\pi]{MHz}$]
 & $P_\mathrm{scatt}$ &  $F$ \\
\colrule
(a) & $\N0.08$ & $\N1.3 \pm 0.1$ & $\N\left(33^{+7}_{-6}\right)\%$ & $0.94$ \\
(b) & $\N0.16$ & $\N1.9 \pm 0.2$ & $\N\left(56^{+8}_{-9}\right)\%$ & $0.95$ \\
(c) & $\N0.48$ & $\N3.2 \pm 0.3$ & $\N\left(91^{+4}_{-6}\right)\%$ & $0.93$ \\
(d) & $ 10.75$ & $ 15.2 \pm 1.5$ & $ \N100\%$ & $0.94$ \\
\end{tabular}
\end{ruledtabular}
\caption{\label{tab:results}%
Experimental settings: power of the laser used to drive the $|0\rangle \leftrightarrow |e\rangle$ transition measured directly before the experiment chamber, corresponding Rabi 
 frequency $\Omega$, photon scattering probability $P_\mathrm{scatt}= 1-\abs{g_0}^2$,
 and experimental process fidelity $F$ 
 with the model process $\Lambda_m$.
The statistical uncertainty of $F$ is $\pm0.01$.}
\end{table}

The fidelity $F$ between the experimental process and the model process $\Lambda_m$ is given by
\begin{equation}
F= \frac19\left[\tr\left(\sqrt{ \sqrt{\chi_\mathrm{m}}
 \chi_\mathrm{exp} \sqrt{\chi_\mathrm{m}}}\right)\right]^2,
\label{eq:fidelity}
\end{equation}
with model Choi matrix $\chi_m$ and experimental Choi matrix $\chi_{\text{exp}}$.

\section{Appendix: Initial state preparation}
\begin{table}[h]
\centering
\begin{ruledtabular}
\begin{tabular}{cll}
$j$ & $U_j$ & $\ket{\psi_j} = U_j\ket0$ \\
\colrule
1 & $\openone$ & $\ket0$ \\
2 & $R^1_y(\pi)$ & $\ket1$ \\
3 & $R^2_y(\pi)$ & $\ket2$ \\
4 & $R^1_y(\pi/2)$ & $\frac{1}{\sqrt{2}}(\ket0+\phantom{i}\ket1)$ \\
5 & $R^1_{-x}(\pi/2)$ & $\frac{1}{\sqrt{2}}(\ket0+i\ket1)$ \\
6 & $R^2_y(\pi/2)$ & $\frac{1}{\sqrt{2}}(\ket0+\phantom{i}\ket2)$ \\
7 & $R^2_{-x}(\pi/2)$ & $\frac{1}{\sqrt{2}}(\ket0+i\ket2)$ \\
8 & $R^2_y(\pi) R^1_y(\pi/2)$ & $\frac{1}{\sqrt{2}}(\ket1+\phantom{i}\ket2)$ \\
9 & $R^2_{y}(\pi) R^1_{-x}(\pi/2)$ & $\frac{1}{\sqrt{2}}(\ket1+i\ket2)$ \\
\end{tabular}
\end{ruledtabular}
\caption{\label{tab:tomography}%
Pulse sequences for the implementation of the unitaries $U_j$, as used for the 
 preparation and measurement of the qutrit.
$R^{\ell}_{\hat{n}}(\phi)$ is a rotation in the qutrit subspace spanned by 
 $\ket0$ and $\ket\ell$ of angel $\phi$ about axis $\hat{n}$.
The pulse sequence is applied from right to left.}
\end{table}

\begin{figure*}[b]
\includegraphics[width=\linewidth]{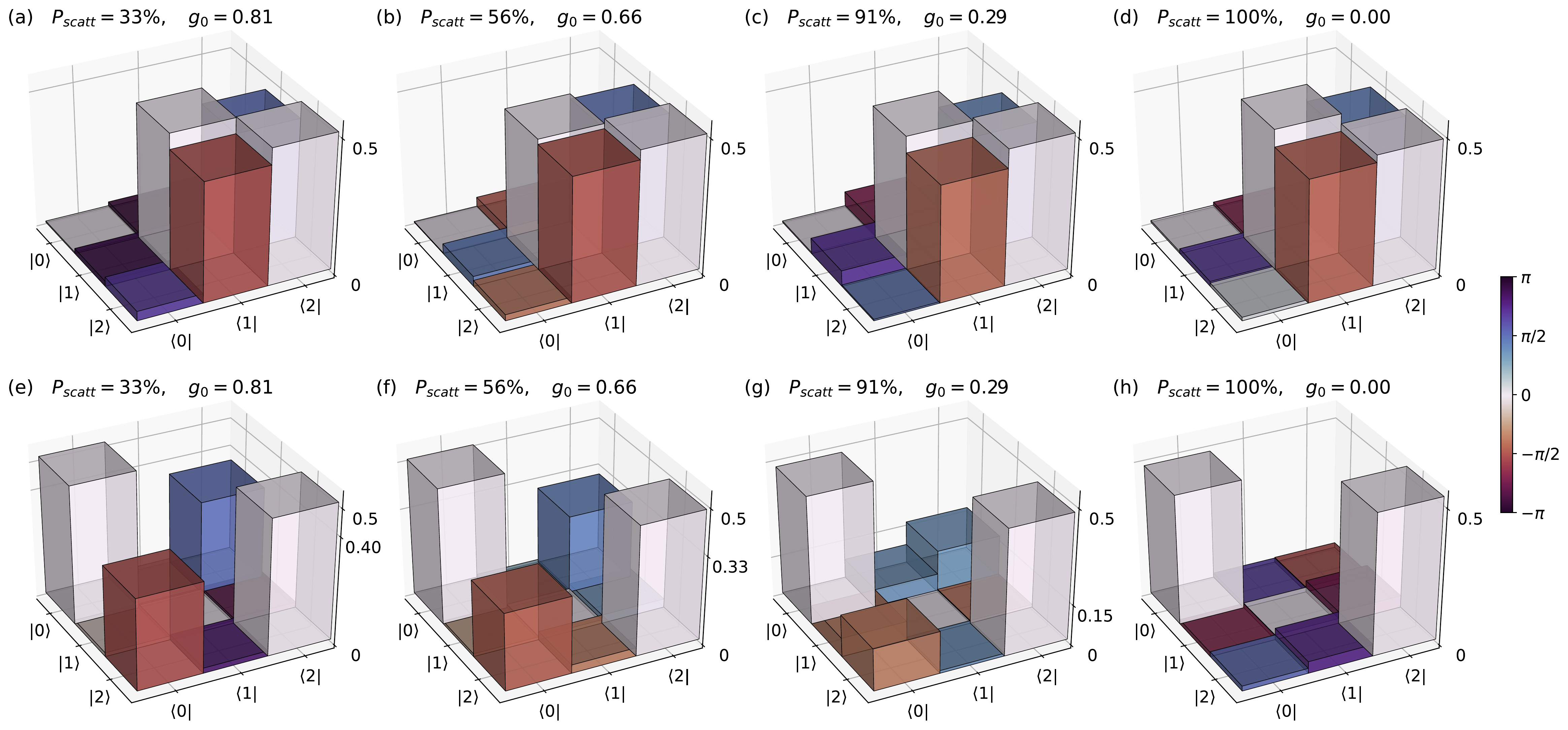}
\caption{\label{fig:rho}%
Reconstructed density matrices $\rho$ from the experimental data for the initial states 
$\frac{1}{\sqrt{2}}\left(\ket{1} + i\ket{2}\right)$ [(a) -- (d)] and $\frac{1}{\sqrt{2}}\left(\ket{0} + i\ket{2}\right)$ [(e) -- (h)]. The colors indicate the complex phase,
$g_0$ is calculated according to our model, see Eq.~(\ref{eq:a0}).
From (a) to (d) and from (e) to (h) the coupling strength of $\ket{0}$ to the photon environment is increased and coherences involving state $\ket{0}$ are destroyed.
Thus the off-diagonal elements are preserved from (a) to (d) (state 
$\frac{1}{\sqrt{2}}\left(\ket{1} + i\ket{2}\right)$) while the off-diagonal elements decrease from
 (e) to (h) (state $\frac{1}{\sqrt{2}}\left(\ket{0} + i\ket{2}\right)$).}
\end{figure*}

\begin{turnpage}
\begin{figure}[h]
\centering
\includegraphics[width=\linewidth]{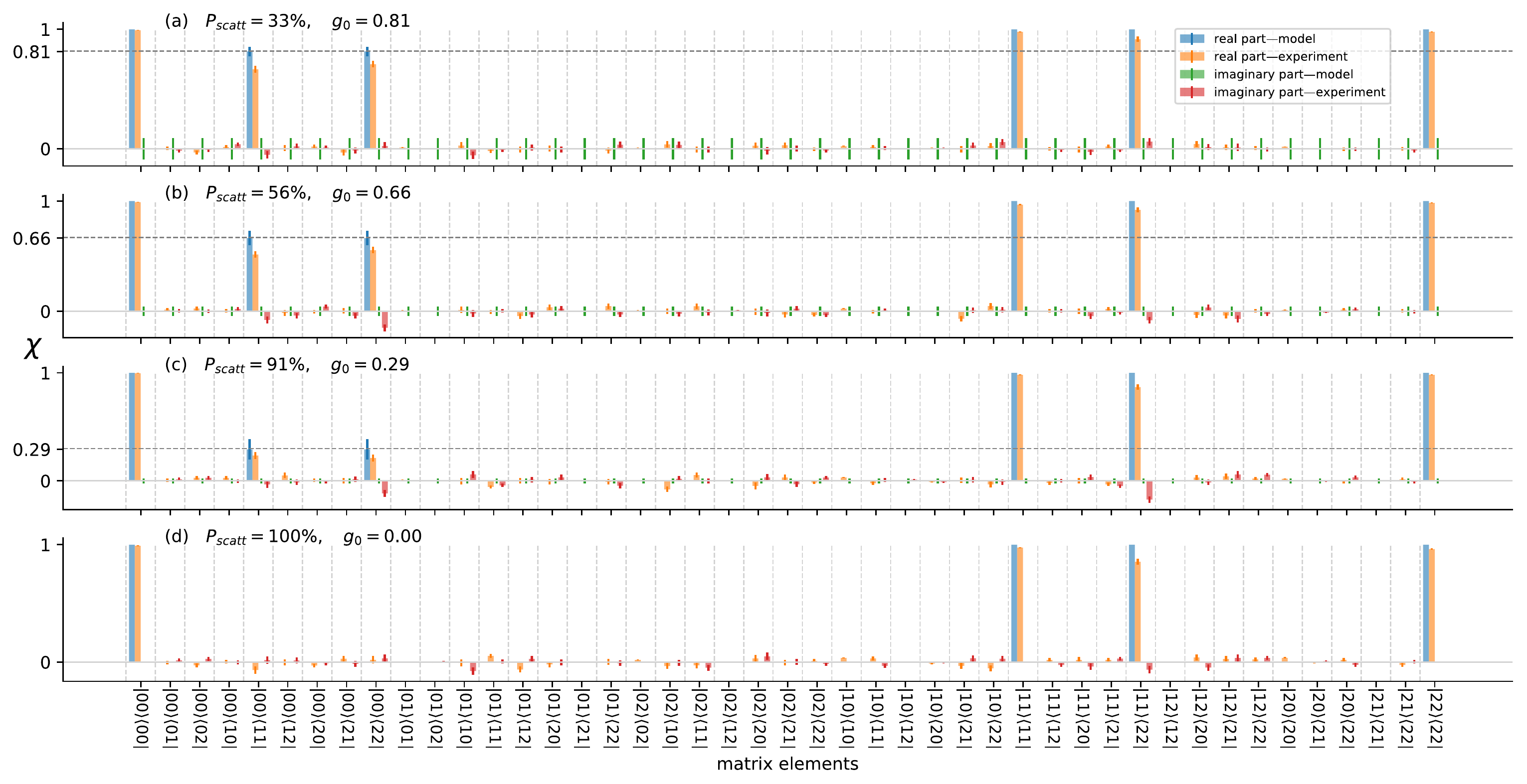}
\label{fig:exp+data}
\caption{Comparison of the elements of the experimentally-determined Choi matrices $\chi_\mathrm{exp}$ and model predictions $\chi_m$.
From (a) to (d) the coupling strength of $\ket{0}$ to the photon environment is increased (see Appendix Table~\ref{tab:results}) and coherences involving state $\ket{0}$ are destroyed.
(d) is most similar to the ideal Lüders process, from (d) $\rightarrow$ (a) the process becomes more similar to the trivial process $\rho \mapsto \rho$.
Error bars in the model result from uncertainties in the experimental parameters $\Omega$ and $\Delta$ (68\% confidence interval).
Error bars in the experimentally-determined Choi matrix elements result from quantum projection noise (68\% confidence interval).}
\end{figure}
\end{turnpage}


\end{document}